\documentclass[prd,superscriptaddress,twocolumn,nofootinbib,showpacs,preprintnumbers,floatfix]{revtex4}

\usepackage{mathrsfs}
\usepackage{amsfonts,amssymb,amsmath}
\usepackage{graphicx,epsfig}
\usepackage{multirow}
\usepackage{verbatim}

\def\fsu5{$\cal{F}$-$SU(5)$}
\def\bfsu5{$\boldsymbol{\mathcal{F}}$-$\boldsymbol{SU(5)}$}

\def\m1half{$M_{1/2}$}
\def\m3half{$M_{3/2}$}
\def\m32{$M_{32}$}

\def\fb{${\rm fb}^{-1}$~}

\def\mt2{$M_{T2}$}
\def\x2{$\chi^2$}
\def\2b{$M_{T2}b$}

\def\bs0{$B_S^0 \rightarrow \mu^+ \mu^-$}

\begin{document}

\title{No-Scale \bfsu5 in the Light of LHC, Planck and XENON}

\author{Tianjun Li}

\affiliation{State Key Laboratory of Theoretical Physics and Kavli Institute for Theoretical Physics China (KITPC),
Institute of Theoretical Physics, Chinese Academy of Sciences, Beijing 100190, P. R. China}

\affiliation{George P. and Cynthia W. Mitchell Institute for Fundamental Physics and Astronomy, Texas A$\&$M University, College Station, TX 77843, USA}

\author{James A. Maxin}

\affiliation{George P. and Cynthia W. Mitchell Institute for Fundamental Physics and Astronomy, Texas A$\&$M University, College Station, TX 77843, USA}

\affiliation{Department of Physics and Astronomy, Ball State University, Muncie, IN 47306 USA}

\author{Dimitri V. Nanopoulos}

\affiliation{George P. and Cynthia W. Mitchell Institute for Fundamental Physics and Astronomy, Texas A$\&$M University, College Station, TX 77843, USA}

\affiliation{Astroparticle Physics Group, Houston Advanced Research Center (HARC), Mitchell Campus, Woodlands, TX 77381, USA}

\affiliation{Academy of Athens, Division of Natural Sciences, 28 Panepistimiou Avenue, Athens 10679, Greece}

\author{Joel W. Walker}

\affiliation{Department of Physics, Sam Houston State University, Huntsville, TX 77341, USA}


\begin{abstract}

We take stock of the No-Scale \fsu5 model's experimental status and prospects
in the light of results from LHC, Planck, and XENON100.  Given that no conclusive
evidence for light Supersymmetry (SUSY) has emerged from the $\sqrt{s} = 7, 8$~TeV collider
searches, the present work is focused on exploring and clarifying the precise nature
of the high-mass cutoff enforced on this model at the point where the
stau and neutralino mass degeneracy becomes so tight that cold dark matter
relic density observations cannot be satisfied.  This hard upper boundary
on the model's mass scale constitutes a top-down theoretical
mandate for a comparatively light (and testable) SUSY spectrum which does not excessively
stress natural resolution of the gauge hierarchy problem.  The overlap
between the resulting model boundaries and the expected sensitivities of
the future 14~TeV LHC and XENON 1-Ton direct detection
SUSY / dark matter experiments is described.

\end{abstract}



\preprint{ACT-4-13, MIFPA-13-17}

\maketitle


\section{Introduction}

We have developed a framework named No-Scale \fsu5~\cite{
Li:2010ws, Li:2010mi,Li:2010uu,Li:2011dw, Li:2011hr, Maxin:2011hy,
Li:2011xu, Li:2011in,Li:2011gh,Li:2011rp,Li:2011fu,Li:2011ex,Li:2011av,
Li:2011ab,Li:2012hm,Li:2012tr,Li:2012ix,Li:2012yd,Li:2012qv,Li:2012jf,Li:2012mr,Li:2013hpa}
that is based upon the tripodal foundations of
the dynamically established boundary conditions
of No-Scale Supergravity, the Flipped $SU(5)$ Grand
Unified Theory (GUT), and a pair of TeV-scale
hypothetical ``{\it flippon}'' vector-like
super-multiplets motivated within local F-theory model
building.  The union of these features has
been demonstrated to naturally resolve a number
of standing theoretical problems, and to compare
favorably with experimental observation of the real world.

The aggressively minimalistic formalism of No-Scale
Supergravity~\cite{Cremmer:1983bf,Ellis:1983sf, Ellis:1983ei, Ellis:1984bm, Lahanas:1986uc}
provides for a deep connection to string theory in the infrared limit,
the natural incorporation of general coordinate invariance (general relativity),
a mechanism for Supersymmetry (SUSY) breaking which preserves a vanishing cosmological constant at the tree level
(facilitating the observed longevity and cosmological flatness of our Universe~\cite{Cremmer:1983bf}),
natural suppression of CP violation and flavor-changing neutral currents, dynamic stabilization
of the compactified spacetime by minimization of the loop-corrected scalar potential, and a dramatically
parsimonious reduction in parameterization freedom.
The split-unification structure of flipped $SU(5)$~\cite{Nanopoulos:2002qk,Barr:1981qv,Derendinger:1983aj,Antoniadis:1987dx}
provides for fundamental GUT scale Higgs representations (not adjoints), natural doublet-triplet
splitting, suppression of dimension-five proton decay~\cite{Antoniadis:1987dx,Harnik:2004yp},
and a two-step see-saw mechanism for neutrino masses~\cite{Ellis:1992nq,Ellis:1993ks}.
Modifications to the one-loop gauge $\beta$-function coefficients $b_i$ induced by inclusion of the
vector-like flippon multiplets create an essential flattening of the $SU(3)$ Renormalization Group Equation (RGE)
running ($b_3 = 0$)~\cite{Li:2010ws}, which translates into a wide separation between the primary
$SU(3)_C \times SU(2)_L$ unification near $10^{16}$~GeV and the secondary $SU(5) \times U(1)_X$ unification
near the Planck mass.  The corresponding baseline extension for logarithmic running of the
No-Scale boundary conditions, especially that imposed ($B_\mu = 0$) on the soft SUSY breaking term $B_\mu$
associated with the Higgs bilinear mass mixing $\mu$, allows sufficient room for natural dynamic
evolution into phenomenologically viable values at the electroweak scale.  Associated flattening
of the color-charged gaugino mass scale generates a stable characteristic mass texture of
$M(\widetilde{t}_1) < M(\widetilde{g}) < M(\widetilde{q})$, featuring a light stop and
gluino that are lighter than all other squarks~\cite{Li:2011ab}.

At this vital juncture in the LHC's history, entering an operational
pause for refitting to enable collisions at the full design energy of 14~TeV,
we take stock of the No-Scale \fsu5 model's full experimental status and prospects.
The first phase of LHC data collection at $\sqrt{s} = 7,8$~TeV
was highly noteworthy for both a discovery (the isolation
of a Standard Model (SM) like Higgs boson at around 125~GeV~\cite{:2012gk,:2012gu,Aaltonen:2012qt})
and a null result (the stubborn persistence of SUSY to unambiguously rise above the noise floor with increasing
luminosity and energy).  Both circumstances have pressed
the standard Constrained Minimal Supersymmetric Standard Model (CMSSM)
and Minimal Supergravity (mSUGRA) parameter spaces to an extreme~\cite{Strumia:2011dv,Baer:2012uya,Buchmueller:2012hv}.
As is natural, the \fsu5 model space has likewise diminished as
it has been probed by incoming collider data~\cite{Li:2013hpa}, although
a wealth of sufficient room to maneuver remains. This result is made all
the more remarkable by the fact that the No-Scale \fsu5 construction
takes the form of a Minimal Parameter Model (MPM), with all essential
experimental characteristics established solely by the universal gaugino
mass boundary $M_{1/2}$, according to which the full SUSY particle
spectrum is proportionally rescaled {\it en masse}.  In particular, the
present work is focused on exploring and clarifying the precise nature
of the high-mass cutoff enforced on the model at the point where the
charged stau and neutral lightest supersymmetric particle (LSP)~\cite{Ellis:1983ew} mass degeneracy
becomes so tight that the Planck~\cite{Ade:2013zuv} and WMAP~\cite{Spergel:2003cb,Spergel:2006hy,Komatsu:2010fb,Hinshaw:2012aka} cold dark matter (CDM)
relic density observations cannot be satisfied.  This hard upper boundary
on the model's leading dimensionful parameter, $M_{1/2}$, constitutes a top-down theoretical
mandate for a comparatively light (and testable) SUSY spectrum which does not excessively
stress natural resolution of the gauge hierarchy problem, and is in itself
a rather distinctive and unique No-Scale \fsu5 characteristic.  The overlap
between the resulting model boundaries and the expected sensitivities of
the future 14~TeV LHC and XENON 1-Ton~\cite{xenon} direct detection
SUSY / dark matter experiments will be described.

\section{The No-Scale \fsu5 Model}

Supersymmetry naturally solves
the gauge hierarchy problem in the SM, and suggests (given $R$ parity conservation)
the LSP as a suitable cold dark matter candidate.
However, since we do not see mass degeneracy of the superpartners,
SUSY must be broken around the TeV scale. In GUTs with
gravity mediated supersymmetry breaking, called
the supergravity models,
we can fully characterize the supersymmetry breaking
soft terms by four universal parameters
(gaugino mass $M_{1/2}$, scalar mass $M_0$, trilinear soft term $A$, and
the low energy ratio of Higgs vacuum expectation values (VEVs) $\tan\beta$),
plus the sign of the Higgs bilinear mass term $\mu$.

No-Scale Supergravity was proposed~\cite{Cremmer:1983bf}
to address the cosmological flatness problem,
as the subset of supergravity models
which satisfy the following three constraints:
i) the vacuum energy vanishes automatically due to the suitable
 K\"ahler potential; ii) at the minimum of the scalar
potential there exist flat directions that leave the
gravitino mass $M_{3/2}$ undetermined; iii) the quantity
${\rm Str} {\cal M}^2$ is zero at the minimum. If the third condition
were not true, large one-loop corrections would force $M_{3/2}$ to be
either identically zero or of the Planck scale. A simple K\"ahler potential that 
satisfies the first two conditions is~\cite{Ellis:1984bm,Cremmer:1983bf}
\begin{eqnarray} 
K &=& -3 {\rm ln}( T+\overline{T}-\sum_i \overline{\Phi}_i
\Phi_i)~,~
\label{NS-Kahler}
\end{eqnarray}
where $T$ is a modulus field and $\Phi_i$ are matter fields.
The third condition is model dependent and can always be satisfied in
principle~\cite{Ferrara:1994kg}.
For the simple K\"ahler potential in Eq.~(\ref{NS-Kahler})
we automatically obtain the No-Scale boundary condition
$M_0=A=B_{\mu}=0$ while $M_{1/2}$ is allowed,
and indeed required for SUSY breaking.
Because the minimum of the electroweak (EW) Higgs potential
$(V_{EW})_{min}$ depends on $M_{3/2}$,  the gravitino mass is 
determined by the equation $d(V_{EW})_{min}/dM_{3/2}=0$.
Thus, the supersymmetry breaking scale is determined
dynamically. No-scale supergravity can be
realized in the compactification of the weakly coupled
heterotic string theory~\cite{Witten:1985xb} and the compactification of
M-theory on $S^1/Z_2$ at the leading order~\cite{Li:1997sk}.

In order to achieve true string-scale gauge coupling unification
while avoiding the Landau pole problem,
we supplement the standard ${\cal F}$-lipped $SU(5)\times U(1)_X$~\cite{Nanopoulos:2002qk,Barr:1981qv,Derendinger:1983aj,Antoniadis:1987dx}
SUSY field content with the following TeV-scale vector-like multiplets (flippons)~\cite{Jiang:2006hf}
\begin{eqnarray}
\hspace{-.3in}
& \left( {XF}_{\mathbf{(10,1)}} \equiv (XQ,XD^c,XN^c),~{\overline{XF}}_{\mathbf{({\overline{10}},-1)}} \right)\, ,&
\nonumber \\
\hspace{-.3in}
& \left( {Xl}_{\mathbf{(1, -5)}},~{\overline{Xl}}_{\mathbf{(1, 5)}}\equiv XE^c \right)\, ,&
\label{z1z2}
\end{eqnarray}
where $XQ$, $XD^c$, $XE^c$, $XN^c$ have the same quantum numbers as the
quark doublet, the right-handed down-type quark, charged lepton, and
neutrino, respectively.
Such kind of models can be realized in ${\cal F}$-ree ${\cal F}$-ermionic string
constructions~\cite{Lopez:1992kg},
and ${\cal F}$-theory model building~\cite{Jiang:2009zza,Jiang:2009za}. Thus, they have been 
dubbed ${\cal F}$-$SU(5)$~\cite{Jiang:2009zza}.

\section{Extending the Wedge of Bare-Minimal Constraints}

\begin{figure*}[htp]
        \centering
        \includegraphics[width=0.85\textwidth]{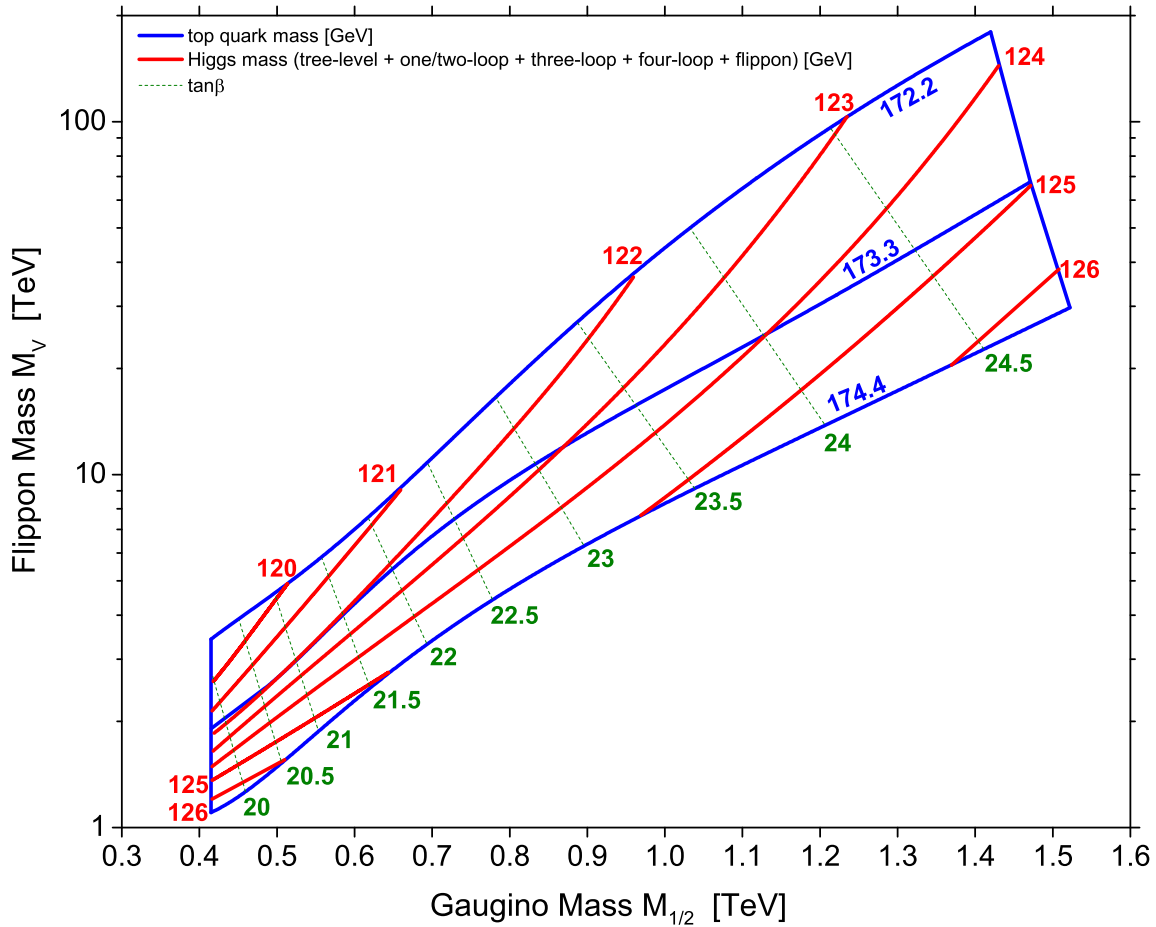}
        \caption{Constrained model space of No-Scale \fsu5 as a function of the gaugino mass $M_{1/2}$ and flippon mass $M_V$. The thick lines demarcate the total Higgs boson mass gradients, including the tree-level plus one/two-loop (as computed by the SuSpect~2.34 codebase), the three-loop plus four-loop contributions, and the flippon contribution. The thin dashed lines represent gradients of tan$\beta$, while the upper and lower exterior boundaries are defined by a top quark mass of $m_t = 173.3 \pm1.1$ GeV. The left edge is marked by the LEP constraints, while the right edge depicts where the Planck relic density can no longer be maintained due to an LSP and light stau mass difference less than the on-shell tau mass. All model space within these boundaries satisfy the Planck relic density constraint $\Omega h^2 = 0.1199 \pm 0.0027$ and the No-Scale requirement $B_{\mu}=0$.}
        \label{fig:higgstanb}
\end{figure*}

\begin{figure*}[htp]
        \centering
        \includegraphics[width=0.85\textwidth]{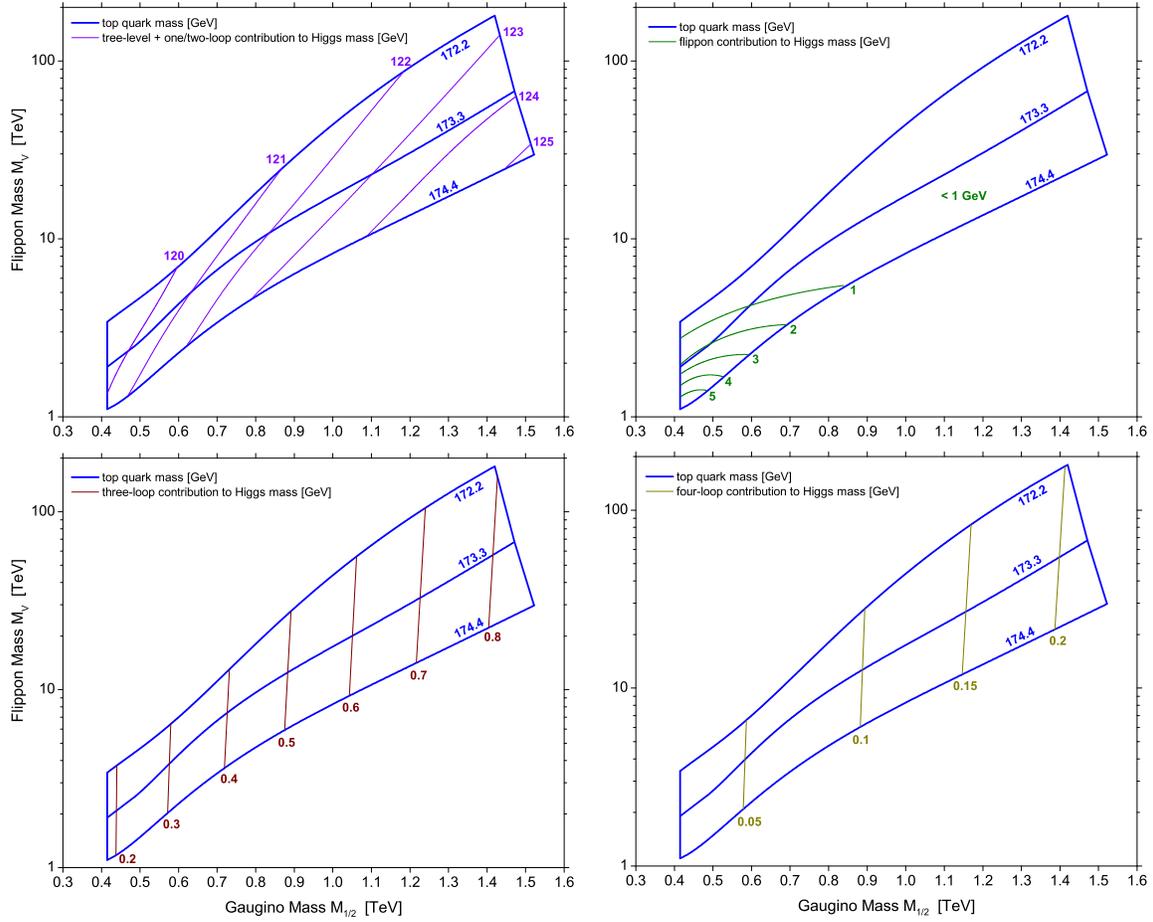}
        \caption{Each panel shows mass gradients within the No-Scale \fsu5 model space of each specific contribution to the total Higgs boson mass illustrated in Fig.~\ref{fig:higgstanb}.}
        \label{fig:higgscomponents}
\end{figure*}

\begin{figure*}[htp]
        \centering
        \includegraphics[width=0.48\textwidth]{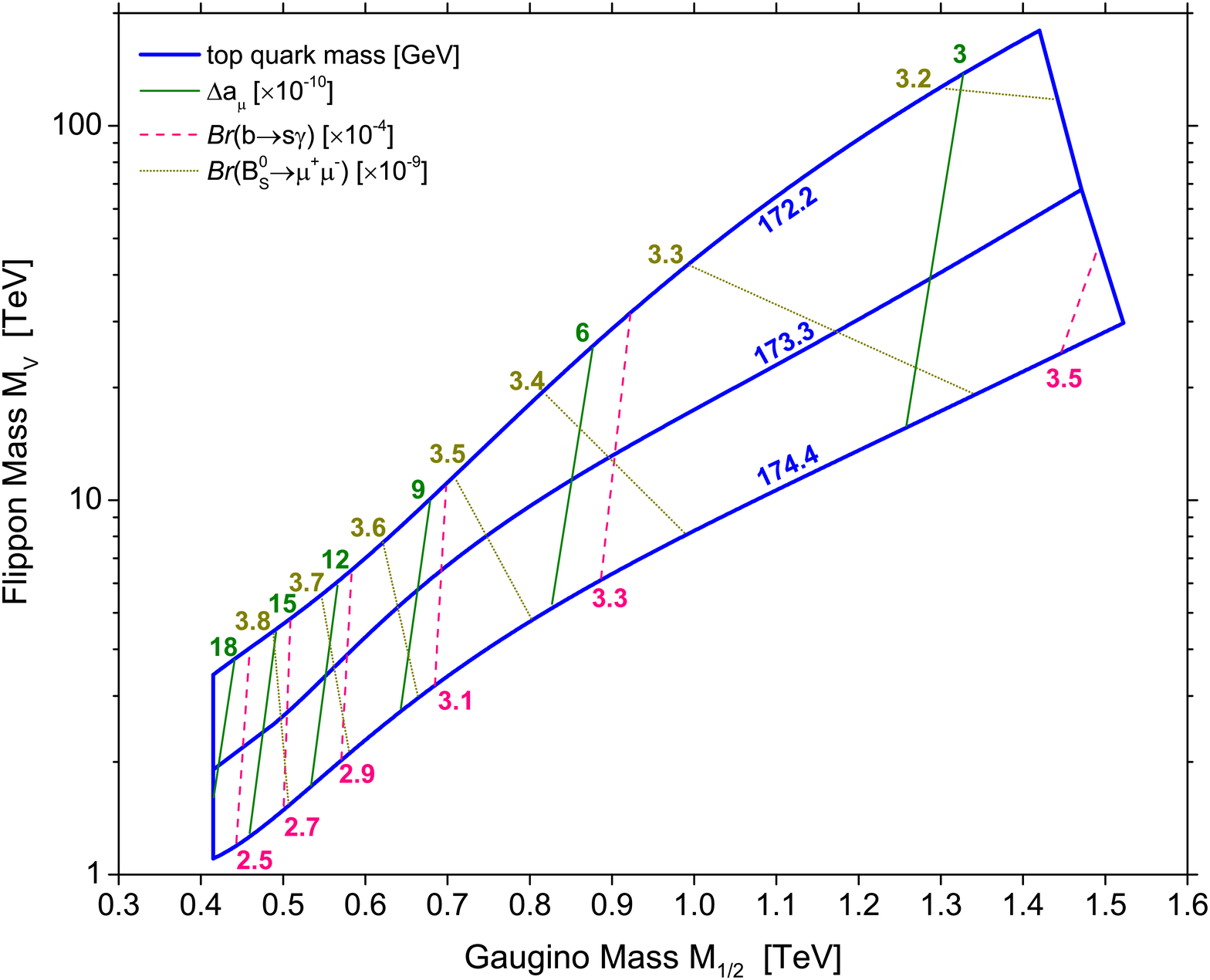}
        \includegraphics[width=0.48\textwidth]{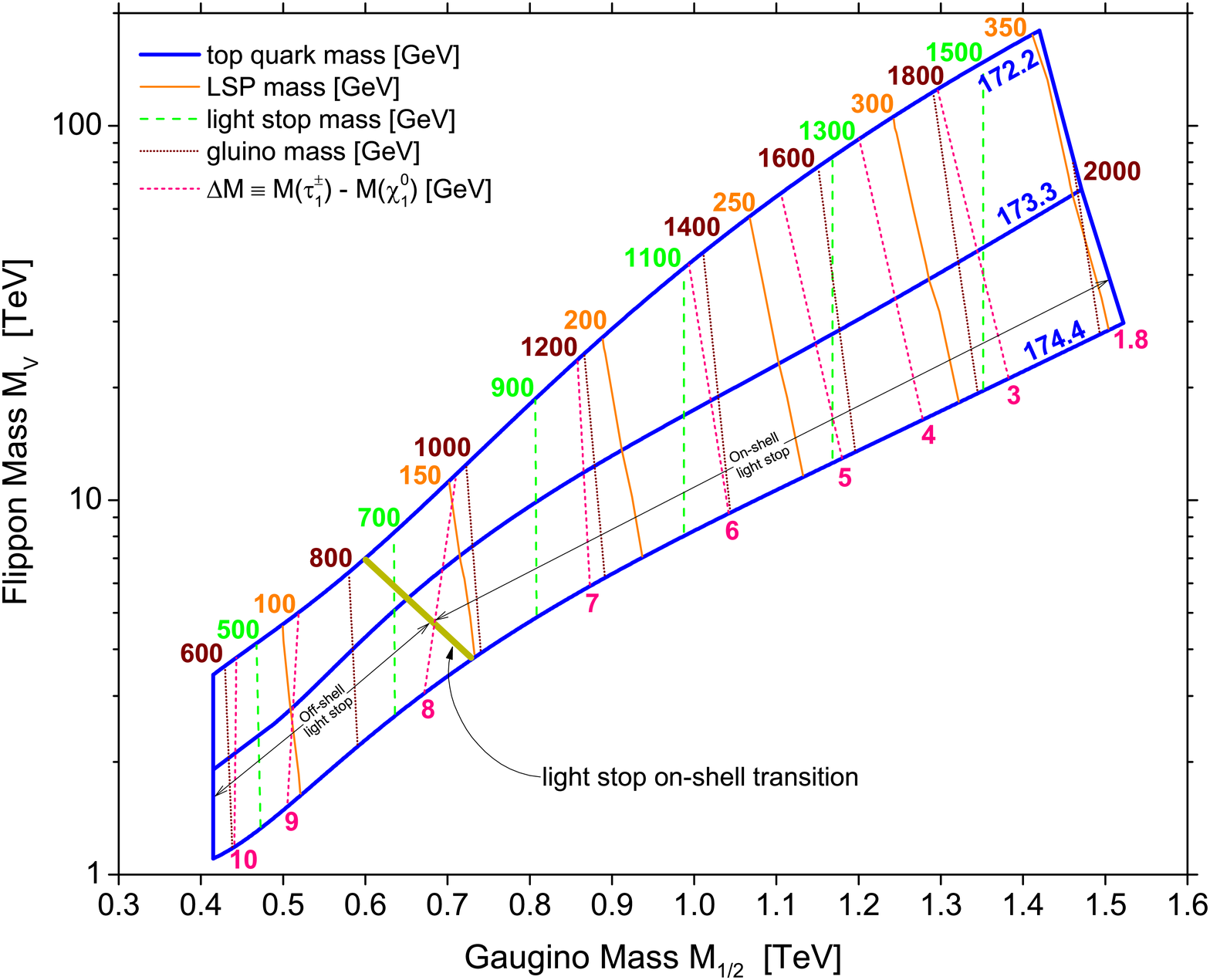}
        \caption{Gradients of the rare-decay process constraints (left) and SUSY masses (right). Also shown are the near degenerate light stau and LSP mass difference, which allows the Planck relic density measurements to be easily satisfied via stau-neutralino coannihilation. The thick line in the right pane illustrates where the gluino-mediated light stop transitions from off-shell to on-shell.}
        \label{fig:rarespectrum}
\end{figure*}

In Ref.~\cite{Li:2011xu}, we presented the wedge of No-Scale \fsu5
model space that is consistent with a set of ``bare minimal''
constraints from theory and phenomenology. The constraints included
i) consistency with the dynamically established boundary
conditions of No-Scale supergravity (most notably the
imposition of a vanishing $B_{\mu}$ at the final flipped $SU(5)$
GUT unification near $M_{\rm Pl}$, enforced as $\left|B_{\mu}\right(M_{\cal F})| \leq 1$ GeV, about
the size of the EW radiative corrections); ii) radiative electroweak
symmetry breaking; iii) the centrally observed WMAP7~\cite{Komatsu:2010fb} CDM relic density (and
now the Planck relic density $\Omega h^2 = 0.1199 \pm 0.0027$~\cite{Ade:2013zuv}) ; iv) the world average
top-quark mass $m_t = 173.3 \pm 1.1$~GeV~\cite{:1900yx}; and  v) precision
LEP constraints on the light SUSY chargino and neutralino mass
content~\cite{LEP}.  This two-dimensional parameterization in the
vector-like {\it flippon} super-multiplet mass scale $M_V$
and the universal gaugino boundary mass scale $M_{1/2}$
was excised from a larger four-dimensional hyper-volume
also including the top quark mass $m_t$ and the ratio
$\tan \beta$. Surviving points, each capable
of maintaining the delicate balance required to satisfy 
$B_\mu = 0$ and the CDM relic density observations, were
identified from an intensive numerical scan, employing
MicrOMEGAs~2.1~\cite{Belanger:2008sj} to compute SUSY masses, using a proprietary modification of the
SuSpect~2.34~\cite{Djouadi:2002ze} codebase to run the
{\it flippon}-enhanced RGEs. The relic density, spin-independent cross-section, and all rare-decay process constraints have been computed with MicrOMEGAs~2.4~\cite{Belanger:2010gh}.

The union of all such points was found to consist of
a diagonal wedge ({\it cf.} Ref.~\cite{Li:2011xu}) in the $M_{1/2}$-$M_V$
plane, the width of which ({\it i.e.} at large $M_{1/2}$ and
small $M_V$ or vice-versa) is bounded by the central 
experimental range of the top quark mass, and the extent
of which ({\it i.e.} at large $M_{1/2} \sim 900$~GeV and large $M_V$) is
bounded by CDM constraints and the transition to a charged
stau LSP.  The advent of substantial LHC collision data in
the SUSY search rapidly eclipsed the tentative low-mass
boundary set by LEP observations.  A substantive correlation
in the \fsu5 mass scale favored by low-statistics excesses
in a wide range of SUSY search channels, particularly lepton-inclusive searches, at both CMS and ATLAS was remarked upon by
our group~\cite{Li:2012mr,Li:2013hpa} just below $M_{1/2} \sim 800$~GeV.
However, a minority of search channels, particularly lepton-exclusive
squark and gluino searches with jets and missing energy~\cite{ATLAS-CONF-2012-109},
were found to yield limits on $M_{1/2}$ that are inconsistent
with this fit, and that exert some limited tension against the upper
$M_{1/2}$ boundary of the model wedge.  This tension is also
reflected in one generic limit of a multijet plus a single lepton SUSY search from the CMS Collaboration
that places the gluino heavier than about 1.3~TeV~\cite{CMS-SUS-13-007}.

If the otherwise successful \fsu5 model phenomenology~\cite{Li:2012mr}
is to be salvaged in the event the limited tension persists against the upper $M_{1/2}$ boundary, it could become imperative to establish a mechanism for either i) reducing the count of events in the offending channels, or ii) extending
the allowed mass scale of the model.  The solution that we
identify, and which is described in the current letter, is
of the second variety.  After extensive study, it became clear that the
upper CDM boundary on the model wedge previously identified at
$M_{1/2} \sim 900$~GeV was premature; in fact, it reflected
not a termination of the ability to provide a viable dark
matter candidate, but rather a termination of the ability
to numerically resolve viable model points within the finite
lattice spacing employed to scan the hyper-volume of parameters. 
Despite this limitation, the scan had provided an essential
foundation for its own extension; given a unique identification
of the preferred top quark mass $m_t$ and ratio $\tan \beta$ for 
each point in the $M_{1/2}$-$M_V$ wedge, the functional form of
this dependency could be numerically extrapolated to continue,
for example, a given string of constant $m_t$ points beyond
the soft $M_{1/2}$ boundary, threading the needle of a steadily
decreasing $m_{{\widetilde{\tau}}_1} - m_{{\widetilde{\chi}}^0_1}$ mass
difference up to the hard limit of the on-shell tau mass at about 1.8~GeV.
Employing this procedure, the wedge was found to be substantially
extended, up to the range of $M_{1/2} \sim 1.5$~TeV.  This region of
the model space corresponds to an exponentially elevated {\it flippon}
mass $M_V$, which may now extend into the vicinity of 100~TeV.  This
delineation of the bare-minimally constrained \fsu5 parameter
space, including the correlated values of $m_t$, $\tan \beta$ and
the light CP-even Higgs mass for each model point, is depicted in
Figure~\ref{fig:higgstanb}.

\begin{figure*}[htp]
        \centering
        \includegraphics[width=1.00\textwidth]{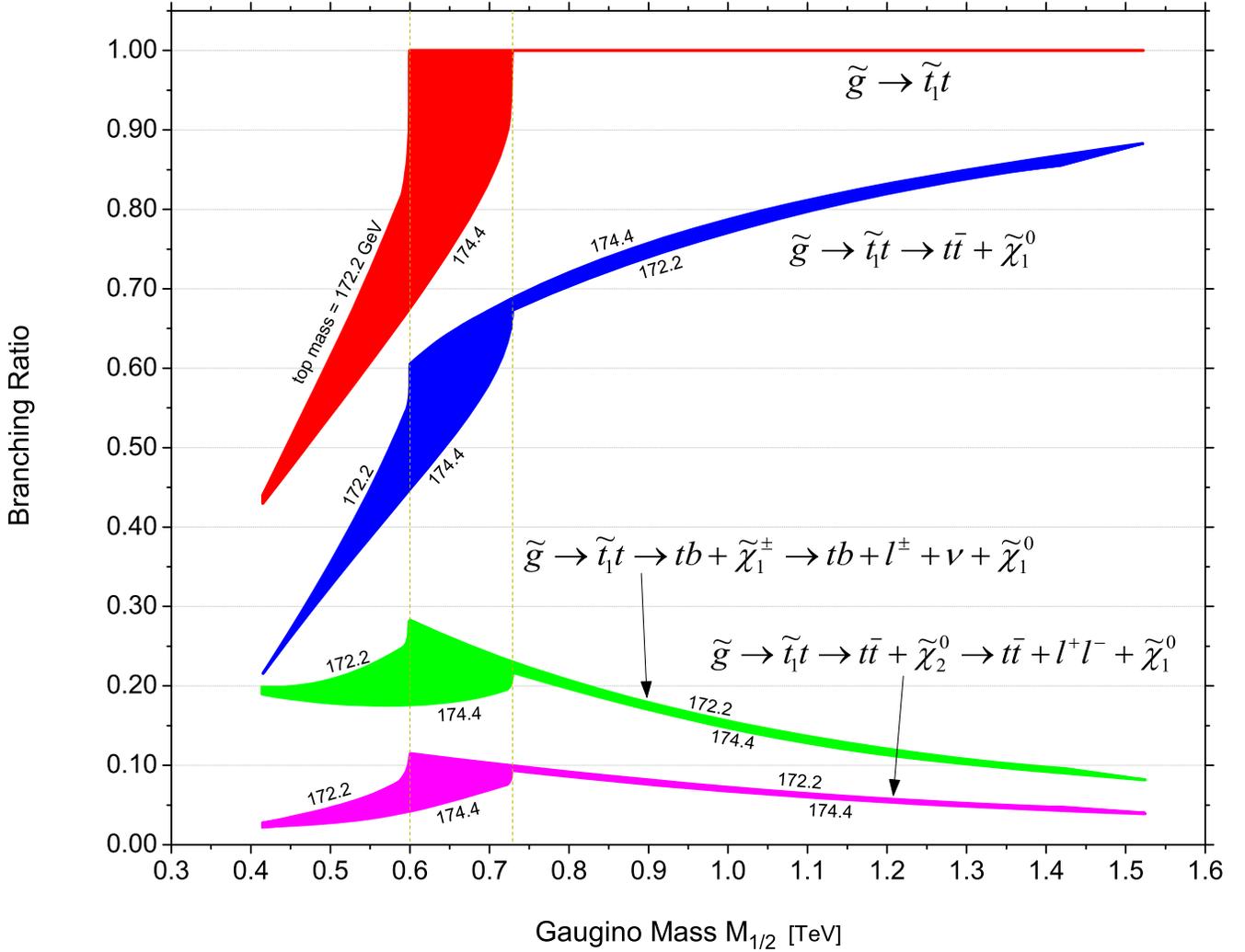}
        \caption{Graphical depiction of the prominent gluino decay channels in No-Scale \fsu5. The branching ratios for each process are shown as a function of the model space depicted in Figs.~(\ref{fig:higgstanb}--\ref{fig:rarespectrum}). The fractions of the lower three regions sum to the total fraction shown for $\widetilde{g} \to \widetilde{t}_1 t$. The top mass borders are shown for orientation against the wedge space shown in Figs.~(\ref{fig:higgstanb}--\ref{fig:rarespectrum}). The vertical dashed lines represent the gluino-mediated light stop off-shell to on-shell transition (where $\widetilde{g} \to \widetilde{t}_1 t$ goes to 100\%) for the two upper and lower extremes of the wedge, as defined by the applied top mass constraint $m_t = 173.3 \pm1.1$ GeV.}
        \label{fig:branchingratio}
\end{figure*}

\begin{figure}[htp]
        \centering
        \includegraphics[width=0.42\textwidth]{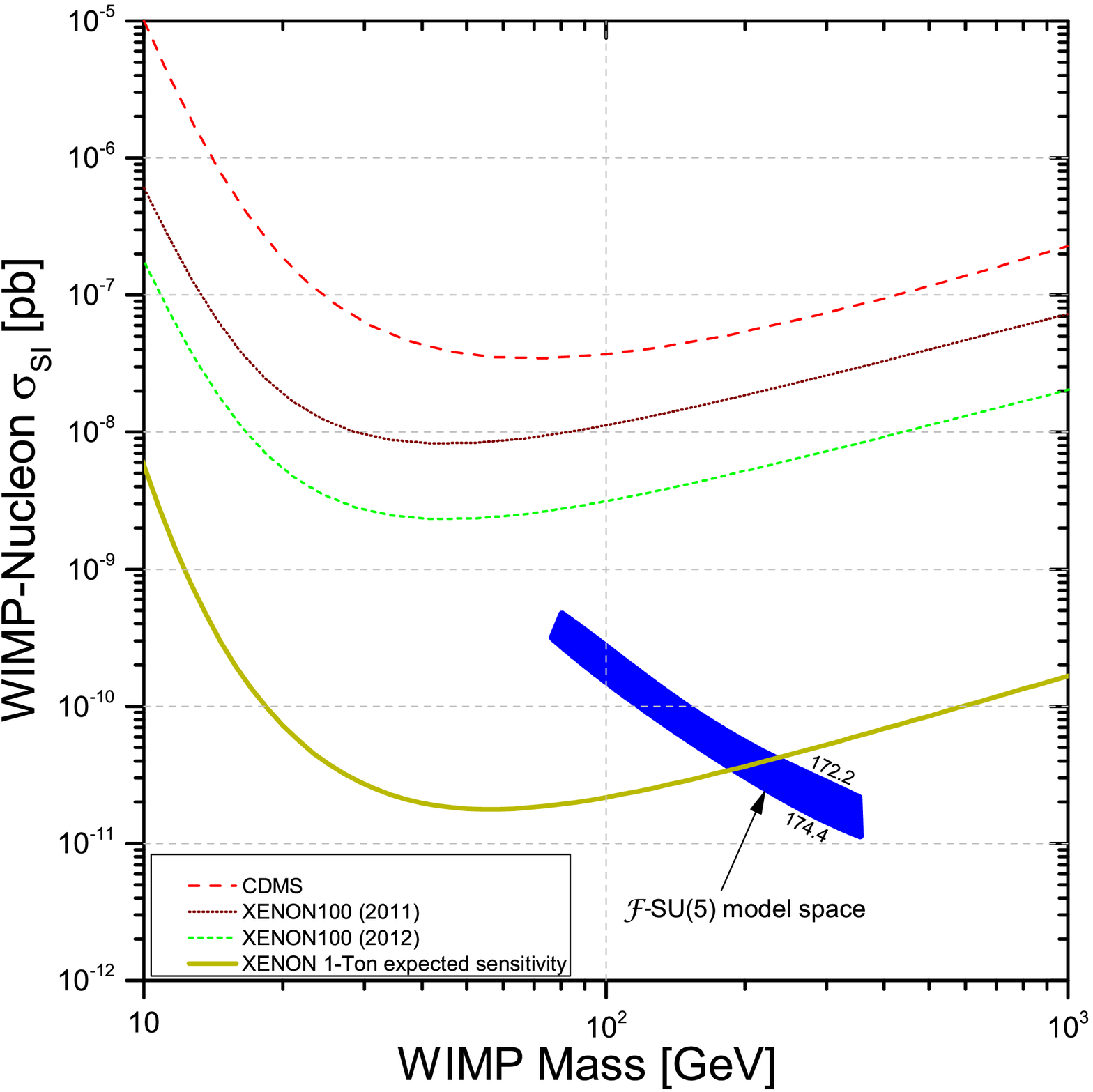}
        \caption{Direct dark matter detection diagram associating the WIMP mass with the spin-independent annihilation cross-section $\sigma_{\rm SI}$. Delineated are the current upper bounds from the CDMS and XENON100 experiments, including the projected sensitivity of the 1-ton XENON experiment (Data courtesy of Elena Aprile and Antonio J. Melgarejo of the XENON Collaboration). The region of parameter space shown is that which is illustrated as a function of $M_{1/2}$ and $M_{V}$ in Figs.~(\ref{fig:higgstanb}--\ref{fig:rarespectrum}).}
        \label{fig:xenon}
\end{figure}

One obvious concern associated with this circumstance is the appearance
of a new intermediate scale of physics, and a potentially new associated
hierarchy problem.  However, we remark that the vector-like {\it flippon}
multiplets are free to develop their own Dirac mass, and are not in
definite {\it a priori} association with the electroweak scale symmetry
breaking; We shall therefore not divert attention in the present work 
to the mechanism of this mass generation, although plausible
candidates do come to mind.

\section{\fsu5 Higgs Phenomenology}

A majority of the $M_{1/2}$ values encompassed by the No-Scale \fsu5
model wedge are consistent with a Higgs boson mass at around 125~GeV,
within the vicinity of the observation reported by CMS, ATLAS,
D\O ~and CDF~\cite{:2012gk,:2012gu,Aaltonen:2012qt}.  This agreement
is generally enhanced at the upper range of the top quark mass limit,
{\it i.e.} for the lower wedge boundary at $m_t = 174.4$~GeV,
as depicted in Figure~\ref{fig:higgstanb}.
The generic conclusion enforced across models that attempt to
cope with the recent Higgs measurement is that either
the stop squark must be rather heavy, around a few TeV, or
there must be non-conventional loop contributions to the Higgs
boson mass, as has been studied previously~\cite{Li:2011ab,Li:2012jf,Li:2012mr}
in the context of our {\it flippon} vector-like supermultiplets.
The specific attribution of various component contributions to the
Higgs mass are summarized in the four panels of Figure~\ref{fig:higgscomponents}.

For smaller $M_{V}$ values (and hence smaller $M_{1/2}$ values in the wedge of \fsu5 model space), the flippon Higgs loops may provide a very large shift to $m_h$, amounting to
2-6~GeV for the lower portion of the model space, although these mass ranges would appear
to be disfavored by the LHC SUSY search.  For intermediate mass ranges up to around 800~GeV, in the vicinity
of the boundary currently probed by collider studies, the {\it flippon} contribution to $m_h$ may still be
appreciable, at around a single GeV.  In the unprobed heavier regions of the model, the correlated increase
in $M_V$ (which is exponential, due to the fact that influence of mass thresholds in the RGEs is logarithmic) means that
these loop contributions to the Higgs mass are suppressed.  However, an interesting complementarity
of the model is revealed in this region, wherein the simultaneously heavier SUSY spectrum contributes
an intrinsic boost to the Higgs mass, compensating for the reduced effect in the vector-like multiplet sector.
In particular, the three- and four-loop MSSM contributions increase as the geometric mean of the two stop masses~\cite{Martin:2007pg},
which is essentially linear in $M_{1/2}$.

It must be emphasized that there are several potential sources of error, which may combine
to provide an uncertainty on the order of 2~GeV in the determination of the \fsu5 Higgs mass, which are
accounted in addition to experimental uncertainties in the measured mass value.  Interestingly, it would seem that the
most notable sources of error may potentially be biased toward underestimation in this case. The tree-level
plus one/two-loop contribution to $m_h$, depicted in the first panel of Figure~\ref{fig:higgscomponents},
are estimated by use of the SuSpect~2.34~\cite{Djouadi:2002ze} algorithm, which publishes an expected error
of $\pm 1.5$~GeV.  By comparison, the FeynHiggs~\cite{Hahn:2010te} package appears systematically predisposed to values
for the bare $m_h$ value that are larger by about the quoted SuSpect error; This leads us to believe that 
corrections to be associated with the SuSpect algorithm may be somewhat more likely positive than negative.
A second important source of error in determination of the Higgs mass is that on the strong coupling $\alpha_s$.
Smaller values of $\alpha_s$ will lower the blue lines (top quark mass contours) of the wedge depicted
in Figure~\ref{fig:higgstanb}, while keeping the red Higgs contours in place, effectively elevating
the range of viable $m_h$ masses.  The Particle Data Group~\cite{pdg2012} world average for $\alpha_s$ is $0.1184 \pm 0.0007$,
giving 2-$\sigma$ boundaries of 0.1170 and 0.1198, and we adopt a value of 0.1172 near to the lower limit.
However, recent extrapolations of $\alpha_s$ from TeV scale physics from the CMS collaboration~\cite{Chatrchyan:2013txa}
and from an outside analysis of ATLAS data~\cite{Malaescu:2012iq} give values of 0.1148 and 0.1151, respectively.
Reanalysis with a coupling closer to this range yields an effective upward shift in the Higgs mass of about 1~GeV~\cite{Li:2012jf}.
Associated theoretical errors are also on the scale of 1-2~GeV, with larger uncertainties associated with 
the unprobed region of the model, at large $M_{1/2}$.

\begin{table*}[htp]
	\centering
	\footnotesize
	\caption{Relic density, spin-independent cross-section, rare-decay process constraints, and supersymmetry and light Higgs boson masses of five representative points, organized in terms of the parameters ($M_{1/2}$, $M_V$, tan$\beta$, $m_t$). All masses specified here are in GeV.}
		\begin{tabular}{|c|c|c|c||c|c|c|c|c||c|c|c|c|c|c|c|c} \hline
$M_{1/2}$&$M_{\rm V}$&$\tan\beta$&$m_{t}$&$\Omega h^2$&$\sigma_{\rm SI}~[{\rm pb}]$&$Br(b \to s\gamma)$&$Br(B_S^0 \to \mu^+ \mu^-)$&$\Delta a_{\mu}$&$m_{\chi^0_1}$&$m_{\widetilde{\tau}_{1}}$&$m_{\chi^{0}_{2},\chi^{\pm}_{1}}$&$m_{\widetilde{t}_{1}}$&$m_{\widetilde{g}}$&$m_{\widetilde{u}_{R}}$&$m_h$ \\ \hline \hline	
$	775	$&$	4800	$&$	22.53	$&$	174.4	$&$	0.1217	$&$	4.9 \times 10^{-11}	$&$	3.22 \times 10^{-4}	$&$	3.48 \times 10^{-9}	$&$	6.5 \times 10^{-10}	$&$	161	$&$	169	$&$	342	$&$	861	$&$	1047	$&$	1475	$&$	124.4	$	\\	\hline

$	990	$&$	8044	$&$	23.34	$&$	174.4	$&$	0.1200	$&$		2.6 \times 10^{-11}		$&$	3.37 \times 10^{-4}	$&$	3.37 \times 10^{-9}	$&$	4.3 \times 10^{-10}	$&$	214	$&$	220	$&$	449	$&$	1104	$&$	1328	$&$	1824	$&$	125.1	$	\\	\hline

$	1200	$&$	30,830	$&$	24.26	$&$	173.3	$&$	0.1218	$&$	2.1 \times 10^{-11}	$&$		3.44 \times 10^{-4}		$&$	3.26 \times 10^{-9}	$&$	3.3 \times 10^{-10}	$&$	276	$&$	281	$&$	572	$&$	1335	$&$	1633	$&$	2102	$&$	124.1	$	\\	\hline

$	1471	$&$	67,690	$&$	24.82	$&$	173.3	$&$	0.1218	$&$		1.5 \times 10^{-11}		$&$	3.50 \times 10^{-4}	$&$	3.20 \times 10^{-9}	$&$	2.3 \times 10^{-10}	$&$	353	$&$	355	$&$	723	$&$	1628	$&$	2010	$&$	2499	$&$	125.0	$	\\	\hline

$	1522	$&$	30,271	$&$	24.73	$&$	174.4	$&$	0.1222	$&$		1.1 \times 10^{-11}		$&$	3.52 \times 10^{-4}	$&$	3.25 \times 10^{-9}	$&$	2.1 \times 10^{-10}	$&$	355	$&$	357	$&$	730	$&$	1684	$&$	2041	$&$	2631	$&$	126.4	$	\\	\hline
		\end{tabular}
		\label{tab:masses}
\end{table*}
\section{Additional Phenomenology and Experimental Prospects}

A {\it Golden} subspace of the original \fsu5 wedge~\cite{Li:2011xu,Li:2013hpa} had been
identified that is further consistent with experimental limits on rare processes,
namely the SUSY contributions to flavor-changing neutral currents $b \to s \gamma$ and
$B^0_s \to \mu^+ \mu^-$, and the anomalous magnetic moment $(g_\mu - 2)/2$ of the muon. 
This golden subspace remains very well represented in the extended wedge.  In general,
the No-Scale \fsu5 model makes rather small contributions to these rare processes, as
is consistent with experimental results. The status of rare process constraints on the \fsu5 model wedge is summarized in the first panel of Figure~\ref{fig:rarespectrum}. 

The most recent world average of \textit{Br}$(b \to s\gamma)$ by the Heavy Flavor Averaging Group (HFAG), BABAR, Belle, and CLEO is $(3.55 \pm 0.24_{\rm exp} \pm 0.09_{\rm model}) \times
10^{-4}$~\cite{Barberio:2007cr}. An alternate tactic to the average~\cite{Artuso:2009jw} yields
a slightly smaller central value, but also a lower error, suggesting
\textit{Br}$(b \to s\gamma) = (3.50 \pm 0.14_{\rm exp} \pm 0.10_{\rm model}) \times 10^{-4}$. See Ref.~\cite{Misiak:2010dz} for recent
discussion and analysis. The theoretical SM contribution at the next-to-next-to-leading order (NNLO)
is estimated at \textit{Br}$(b \to s \gamma) = (3.15 \pm 0.23) \times 10^{-4}$~\cite{Misiak:2006zs} and
\textit{Br}$(b \to s \gamma) = (2.98 \pm 0.26) \times 10^{-4}$~\cite{Becher:2006pu}. The addition of
these errors in quadrature provides the $2\sigma$ limits of
$2.86 \times 10^{-4} \le Br(b \to s \gamma) \le 4.24 \times 10^{-4}$. Even though more higher order SM contributions may be calculated in the future, we consider these constraints to be reasonably stable and to remain so. The present limits would insert a lower limit in \fsu5 at about $M_{1/2} \ge 550$ GeV, well below the constraints set by the 7 and 8 TeV LHC SUSY search data~\cite{Li:2013hpa}. With \textit{Br}$(b \to s\gamma) \sim 3.5 \times 10^{-4}$ at the upper limit of the model space near $M_{1/2} \sim 1500$ GeV, we expect little to no impact on the \fsu5 model space from future measurements of \textit{Br}$(b \to s\gamma)$.

A potentially significant recent observation involves the LHCb~\cite{Aaij:2013aka} and CMS~\cite{Chatrchyan:2013bka} \textit{Br}(\bs0) measurements. The search for this rare decay using 2.0 \fb at 8 TeV and 1.0 \fb at 7 TeV by LHCb and 20 \fb at 8 TeV and 5 \fb at 7 TeV by CMS has observed the first evidence of the process \bs0, where a 4.0--4.3$\sigma$ excess above background expectations was recorded. The results give a branching fraction of $1.9 \times 10^{-9} \le$ \textit{Br}(\bs0) $\le 4.0 \times 10^{-9}$.  This process has been evaluated within the No-Scale \fsu5 viable model space~\cite{Li:2012yd}, computing the parameter space constraints $3.4 \times 10^{-9} \le$ Br(\bs0) $\le 4.0 \times 10^{-9}$ for $400 \le M_{1/2} \le 900$ GeV. Here we extend the analysis to the new upper limit of the model space, giving a lower bound of \textit{Br}(\bs0) $\sim 3.2 \times 10^{-9}$ near $M_{1/2} \sim 1500$ GeV, which is very close to the expected SM branching faction. This is a consequence of the particularly small globally permitted range of $19.4 \lesssim {\rm tan}\beta \lesssim 25$, as shown in Figure~\ref{fig:higgstanb}. As a result, the \fsu5 SUSY contribution to \textit{Br}(\bs0), which is proportional to the sixth power of tan$\beta$, is much smaller than the anticipated SM effect. The LHCb and CMS observations are indeed consistent with the branching ratio expected within the SM, which should constrain or invalidate even more SUSY models, nonetheless, the case for \fsu5 is actually strengthened by the recent measurements. Since the SUSY contribution to \textit{Br}(\bs0) in an \fsu5 framework is quite small for $M_{1/2} \gtrsim 700$ GeV, it will be essentially indistinguishable from the SM only value. Hence, while future measurements of increased precision could inflict more damage upon the global landscape of SUSY models, it is highly improbable that the \fsu5 model space for $M_{1/2} \gtrsim 700$ could be further constrained via \textit{Br}(\bs0).

The anomalous magnetic moment $a_{\mu} \equiv {\rm (g_{\mu}-2)/2}$ of the muon in the SM can be segregated into the electromagnetic, hadronic, and electroweak contributions. The hadronic contribution generates the largest theoretical uncertainty, where its accurate evaluations have to rely on the following experimental measurements:
(1) $e^{+} e^{-} \to {\rm hadrons}$ and $e^{+} e^{-} \to \gamma {\rm hadrons}$;
(2) $\tau^{\pm} \to \nu \pi^{\pm} \pi^0$.
Utilizing the electron data, we get
$\Delta a_{\mu} \equiv a_{\mu}({\rm exp})-a_{\mu}({\rm SM})$ is $28.7 \pm 8.0 \times 10^{-10} $~\cite{Davier:2010nc} and $26.1 \pm 8.0 \times 10^{-10}$~\cite{Hagiwara:2011af}, which correspond respectively to $3.6 \sigma$ and $3.3 \sigma$ discrepancies.
The complete tenth-order QED contributions and improved eighth-order QED contributions were obtained recently, giving $\Delta a_{\mu} = 24.9 \pm 8.7 \times 10^{-10}$~\cite{Aoyama:2012wk}, corresponding to a $2.9 \sigma$ discrepancy.
Using the $\tau$ data, we have
$\Delta a_{\mu} = 19.5 \pm 8.3 \times 10^{-10}$~\cite{Davier:2010nc}, corresponding to a $2.4 \sigma$ discrepancy.
Consequently, the average is $\Delta a_{\mu} = 22.5 \pm 8.5 \times 10^{-10}$ at the 1$\sigma$ deviation.
The $2 \sigma$ deviation of $\Delta a_{\mu} = 22.5 \pm 17 \times 10^{-10}$ to this experimental central value could provide a tentative upper limit in the No-Scale \fsu5 parameter space of $M_{1/2} \simeq 850$ GeV. While we have greater confidence in the pertinence of the limits imposed by the process $b \to s\gamma$ and \bs0, we extend somewhat less credulity to the $a_{\mu}$ measurements and computations. Interestingly, a recent analysis~\cite{Arbuzov:2013mta}
of non-perturbative SM contributions to the muon anomaly concludes that all observed
excesses may be potentially accounted for with no resort whatsoever to SUSY loops. 
Therefore, it is not difficult to argue for a $\Delta a_{\mu}$ value close to zero, again substantially expanding the favored window. If at some point in the future the lower boundary for $\Delta a_{\mu}$ approaches zero or vanishes entirely, then this very tentative upper limit of $M_{1/2} \simeq 850$ GeV will increase or vanish altogether as well.  Likewise, if the anomalous magnetic moment of the muon cannot be constrained any further than $\Delta a_{\mu} = 22.5 \pm 17 \times 10^{-10}$ at 2$\sigma$ in possible forthcoming measurements of higher precision, then it may be impractical to preserve an upper limit in the model space from this measurement.

The spectral composition of the \fsu5 wedge, including the neutralino (Bino) LSP ${\widetilde{\chi}}_1^0$,
light stop ${\widetilde{t}}_1$, and gluino ${\widetilde{g}}$ masses, as well as the stau ${\widetilde{\tau}_1^\pm}$
minus neutralino mass difference, are depicted in the second panel of Figure~\ref{fig:rarespectrum}.
We emphasize that the light stop $\widetilde{t}_1$ and gluino $\widetilde{g}$ are lighter than the bottom squarks $\widetilde{b}_1$ and $\widetilde{b}_2$, top squark $\widetilde{t}_2$, and the first and second
generation left- and right-handed heavy squarks $\widetilde{q}_R$ and $\widetilde{q}_L$ throughout the model wedge, generating a uniquely distinctive test signature at the LHC. This spectrum produces a characteristic event topology starting with the pair production of heavy first or second generation squarks $\widetilde{q}$ and/or gluinos $\widetilde{g}$ in the initial hard scattering process, with each heavy squark likely to yield a quark-gluino pair $\widetilde{q} \rightarrow q \widetilde{g}$ in the cascade decay. The gluino
mediated light stops will be off-shell for gaugino masses below $M_{1/2} \sim 600-700$~GeV, depending on
the value of $M_V$, as depicted by the gold boundary in the mass spectrum plot. In No-Scale \fsu5, 
the gluino proceeds via $\widetilde{g} \rightarrow \widetilde{t}_1 \overline{t}$, where both the on- and off-shell light stops decay as $\widetilde{t}_1 \rightarrow t \widetilde{\chi}_1^0$, $\widetilde{t}_1 \rightarrow b \widetilde{\chi}_1^{\pm}$, or $\widetilde{t}_1 \rightarrow t \widetilde{\chi}_2^0$. The branching ratios for these gluino and light stop channels are shown in Fig.~\ref{fig:branchingratio} as a function of $M_{1/2}$ in the wedge of model space. Pushing to the upper $M_{1/2}$ limit at around 1.5 TeV, where the stau-neutralino mass difference at 1.8~GeV is
almost equal to the tau mass, the heavy squark is about 2.7~TeV, gluino about 2~TeV, light stop about 1.6~TeV,
and LSP neutralino about 350~GeV.  Given the strong light stau and LSP neutralino mass degeneracy in this portion
of the model, one may make an additional interesting prediction for LHC phenomenology: in stau production,
the tau and LSP neutralino missing momentum signal will be collinear.

In Figure~\ref{fig:xenon}, we conclude our analysis with a depiction of the reach of the current CDMS~\cite{Ahmed:2008eu,Ahmed:2009zw,Ahmed:2011gh} and XENON100~\cite{Aprile:2010um,Aprile:2011hx,Aprile:2011hi,Aprile:2012nq} dark matter direct detection experiments (neither of which can probe any portion of the \fsu5 wedge), and a projection of the expected sensitivity of the future 1-Ton XENON experiment~\cite{xenon}. This latter experiment is expected to probe the \fsu5 model wedge up to around a WIMP mass of 200~GeV, corresponding roughly to $M_{1/2} \sim 900$~GeV.
This is a scale comparable to the existing reach of the 7 and 8~TeV collider searches, but well behind the anticipated reach of the 14~TeV LHC into the No-Scale \fsu5 model space~\cite{LMNW-P}.

\section{Conclusions}

We have considered the No-Scale \fsu5 model's experimental status and prospects
in the light of results from LHC, Planck, and XENON100.  Given that no conclusive
evidence for light Supersymmetry has emerged from the $\sqrt{s} = 7, 8$~TeV collider
searches as of yet, this work focused on exploring and clarifying the precise nature
of the high-mass cutoff enforced on this model at the point where the
stau and neutralino mass degeneracy becomes so tight that cold dark matter
relic density observations cannot be satisfied.  This hard upper boundary
on the model's mass scale constitutes a top-down theoretical
mandate for a comparatively light (and testable) SUSY spectrum which does not excessively
stress natural resolution of the gauge hierarchy problem.  The overlap
between the resulting model boundaries and the expected sensitivities of
the future 14~TeV LHC and XENON 1-Ton direct detection
SUSY / dark matter experiments were described.  Coverage up to a WIMP mass around 200~GeV is expected in the latter
case, corresponding to roughly the lower half of the \fsu5 model wedge, by the $M_{1/2}$
gaugino mass scale.

\begin{acknowledgments}
This research was
supported in part
by the DOE grant DE-FG03-95-Er-40917 (TL and DVN),
by the Natural Science Foundation of China
under grant numbers 10821504, 11075194, 11135003, and 11275246 (TL),
and by the Mitchell-Heep Chair in High Energy Physics (JAM).
We also thank Sam Houston State University
for providing high performance computing resources.
\end{acknowledgments}


\bibliography{bibliography}

\end{document}